\documentclass[aps,pre,twocolumn,superscriptaddress,groupedaddress]{revtex4}  % for review and submission

\usepackage{graphicx}  % needed for figures
\usepackage{dcolumn}   % needed for some tables
\usepackage{bm}        % for math
\usepackage{amssymb}   % for math
\usepackage{amsmath}
\usepackage{color,soul}

\begin{document}

% The following information is for internal review, please remove them for submission
\widetext
%\leftline{Version xx as of \today}
%\leftline{Primary authors: Joe E. Physics}
%\leftline{To be submitted to (PRL, PRD-RC, PRD, PLB; choose one.)}
%\leftline{Comment to {\tt d0-run2eb-nnn@fnal.gov} by xxx, yyy}
%\centerline{\em D\O\ INTERNAL DOCUMENT -- NOT FOR PUBLIC DISTRIBUTION}

% the following line is for submission, including submission to the arXiv!!
%\hspace{5.2in} \mbox{Fermilab-Pub-04/xxx-E}

\title{Information-theoretic measures for
non-linear causality detection:  application to social media sentiment and cryptocurrency prices}
\affiliation{Department of Computer Science \& Centre for Blockchain Technologies, University College London, Gower Street,  WC1E 6EA, London, United Kingdom}

\author{Z.~Keskin} \affiliation{Department of Computer Science \& Centre for Blockchain Technologies, University College London, Gower Street, WC1E 6EA, London, United Kingdom}
\affiliation{Department of Physics and Astronomy, University College London, Gower Street, WC1E 6EA, London, United Kingdom}
\author{T.~Aste} \affiliation{Department of Computer Science \& Centre for Blockchain Technologies, University College London, Gower Street,  WC1E 6EA, London, United Kingdom}

% visitor_addresses.tex                       18 June 2018
%  available symbols are:
%  $\ast, \dag, \ddag, \S, \P, $\|$, $\ast\ast$, \dag\dag, \ddag\ddag ,\#
%
%
\vskip 0.25cm
                             % of this file prior to submission, they
                             % contain a time stamp for the authorlist)
                             % (includes institutions and visitors)
\date{\today}

\begin{abstract}
  Information transfer between time series is calculated by using the asymmetric information-theoretic measure known as transfer entropy. Geweke's autoregressive formulation of Granger causality is used to find linear transfer entropy, and Schreiber's general, non-parametric, information-theoretic formulation is used to detect non-linear transfer entropy. 
  \noindent
  We first validate these measures against synthetic data. Then we apply these measures to detect causality between social sentiment and cryptocurrency prices. We perform significance tests by comparing the information transfer against a null hypothesis, determined via shuffled time series, and calculate the Z-score. We also investigate different approaches for partitioning in nonparametric density estimation which can improve the significance of results.
  \noindent
  Using these techniques on sentiment and price data over a 48-month period to August 2018, for four major cryptocurrencies, namely bitcoin (BTC), ripple (XRP), litecoin (LTC) and ethereum (ETH), we detect significant information transfer, on hourly timescales, in directions of both sentiment to price and of price to sentiment. We report the scale of non-linear causality to be an order of magnitude greater than linear causality.
  \noindent
  
\end{abstract}

%\pacs{}
\maketitle

\section{Introduction}

  Causality is {\color{black} a central concept in natural sciences, commonly understood to describe} where some process, evolving in time, has some observable effect on a second process. However, the nature of this causative effect is challenging to describe and quantify with precision. There is a long history in determining whether some change truly causes another \cite{hume1738treatise,pearl2009causality}, especially if the effect is not deterministic, and is observed only in aggregate. In this paper, we consider a statistical form of causality, which can be observed in co-dependent time series where a response in the dependent series is more likely to follow after some change in the driving series. The direction of information transfer is forced by requiring the cause to precede the effect. This concept was conceived first by Wiener in 1956 \cite{Wiener}, and formalised by Granger in 1969 \cite{granger1969investigating} who was subsequently awarded the Nobel memorial prize in economics for his work on the analysis of time series. In simplest terms, the so-called Granger causality describes by how much a response in the dependent series can be explained by a change in the first; or, more exactly, the extent to which a given series is better able to be predicted by considering the information provided by a prior sequence of another series. If this response scales as a linear multiple of the driving signal, this relationship is described as a linear coupling. If, instead, the response follows some other function of the signal, the relationship is non-linear.

  In modern portfolio theory, investors commonly calculate correlations between asset types to construct portfolios aiming to maximise their return at a given level of risk \cite{markowitz1952portfolio}. In the search for excess returns, quantitative approaches are often exploited to detect predictive signals across time series. In the ideal case, from knowing the movement of one price, we can infer the movement of a second. For an investor, it is sufficient to know that the first movement anticipates the second, and in this paper we explore the effectiveness of two promising techniques for detecting anticipatory signals between alternative data and cryptocurrency prices. 

  The concept of an entirely peer-to-peer digital currency managed via a distributed ledger was described and applied by Nakamoto in 2008 \cite{nakamoto2008bitcoin}, who named the currency `Bitcoin'. The proposal and subsequent implementation captured the attention of technologists, economists, libertarians and futurists, and spawned numerous adaptations utilising the blockchain technology \cite{aste2017blockchain}, which have come to be known as cryptocurrencies. Trading in these cryptocurrencies has become widely available even to less sophisticated retail investors, and volumes have grown significantly as interest in the currencies has widened. The crypto market is characterised by high volatility which seems to reflect changes in the attitudes of investors.  {\color{black} The usage of cryptocurrencies in the traditional economy remains limited, and it is reasonable to assume that prices are in part driven by speculative dynamics, separate to any utility as a medium of exchange or to any revenue-generating process}. Therefore a similar, and more marked predictive effect as observed in equity markets, should be observed between measures of social media market sentiment and cryptocurrency prices.
We therefore hypothesise that investor sentiment on future prices may be expected to feed into the short-term price movements via speculation. 
This paper tests this hypothesis.

{\color{black} The relationship between social media sentiment and price has been explored in the literature for traditional markets and, more recently, for crypto markets as well.} For instance, Bollen et al.  \cite{bollen2011twitter} showed that the mood of Twitter messages can be used as a proxy for market sentiment, and that this can show a linear relationship with price movements in US equities. Zheludev \& Aste \cite{zheludev2014can} also performed sentiment analysis using Natural Language Processing (NLP) on Twitter data, to show sentiment is significantly coupled with price movements for a number of instruments issued by S\&P500 firms. Souza \& Aste  used Twitter messages to model market sentiment, and showed the non-linear predictive relationship may be greater than the linear one \cite{souza2016nonlinear}. 
  
  In the cryptocurrency market specifically, one of the authors of the present paper has recently applied information-theoretic techniques to approaches from network theory to characterise the structure of the market as a complex system \cite{Aste2019}. This provided evidence that the market forms a complex, causally-interrelated network linking prices and sentiments across multiple currencies.

  Hypothesising, therefore, that cryptocurrency price depends on prior values of both price and market sentiment, a Granger causality test can detect the impact of past values of $X_t$ on future values of $Y_t$ \cite{granger1969investigating}. This can be calculated using a vector auto-regressive (VAR) model, which describes the extent to which including past values of $X$, at some time-lag $k$, reduces the sum of squared residuals in the regression of $X$ against $Y$, hence estimating the predictive effect of the social sentiment at time $t-k$ on the price at time $t$. 

  The VAR approach performs a regression analysis which is limited to linear associations between variables. To investigate non-linear effects, we can adopt techniques developed in information theory. Many popular information-theoretic measures for comparing distributions, such as mutual information, are symmetric and so can not model a directional information transfer from $X$ to $Y$. Therefore, to generalise Granger causality to the non-linear case, we adopt the measure formalised by Schreiber \cite{schreiber2000measuring}, known as transfer entropy, which is able to capture the size and also the direction of information transfer.

  Transfer entropy arises from the formulation of conditional mutual information; when conditioning on past values of the variables, it  quantifies the reduction in uncertainty provided by these past values in predicting the dependent variable. This presents a natural way to model statistical causality between variables in multivariate distributions. In the general formulation, transfer entropy is a model-free statistic, able to measure the time-directed transfer of information between stochastic variables, and therefore provides an asymmetric method to measure information transfer. As presented in this paper, transfer entropy appears naturally as a generalisation of Granger causality. In fact it has been shown that, for multivariate normally-distributed statistics, where the relationship is therefore linear, this is indeed the case; Granger causality and transfer entropy are equivalent \cite{barnett2009granger}.

  Though developed relatively recently, information-theoretic methods have been used with success in research across disciplines, to detect information transfer where interventionist approaches are not possible. For example, in neuroscience, Vicente et al. \cite{vicente2011transfer} found transfer entropy to be a superior measure in detecting causality in electrophysiological communication than the auto-regressive Granger causality formulation. In climatology, Liang  derived from first principles a linear information flow measure, and used this to show that El Ni\~{n}o tends to stabilise the Indian Ocean Dipole \cite{san2014unraveling}. This analysis also detected a causal effect in the other direction; the Indian Ocean Dipole was shown to amplify El Ni\~{n}o oscillations. The technique was used with further success by Stips et al. \cite{stips2016causal} to confirm that recent CO$_2$ emissions show a one-way causality towards global mean temperature anomalies, but that on paleoclimate timescales, this direction is reversed and temperatures drive CO$_2$ levels. Finally, in finance, information transfer was measured between equities indices by Kwon \& Yang, showing that the information transfer was greatest from the US, and greatest towards the APAC region \cite{kwon2008information}. In particular the S\&P500 was shown to be the strongest driver of other stock indices. In an earlier and somewhat related work, Marschinski \& Kantz   \cite{Marschinski2002} defined and used effective transfer entropy to quantify contagion in financial markets. Similarly, Tungsong et al. \cite{tungsong2018} developed upon the previous work by Diebold \& Yilmaz \cite{diebold2009measuring} in quantifying spillover effects between financial markets, generalising the methodology and estimating the time evolution of interconnectedness between financial systems.

 The rest of the paper is organised as follows. In Section \ref{s.background} we provide a brief background on Granger causality (linear causality measure) and transfer entropy (non-linear causality measure).  In Section \ref{s.method} we describe details of the methodology adopted to quantify and validate linear and non-linear causality, and the techniques used to generate synthetic series of linear and non-linear causal coupling. Section \ref{s.validation} demonstrates that the methodologies correctly detect causality  in the linear and non linear case when testing against synthetic data. Results for real data, concerning causality between cryptocurrency price and sentiment, are presented in Section \ref{s.results}. Section \ref{s.conclusions} reports conclusions and perspectives.

\section{Background} \label{s.background}  

  We calculate statistical causality between time series using two different approaches. The first assumes linearity and employs vector auto-regressive techniques to estimate the extent to which knowing the driving time series can {\color{black} help predict} the dependent series. 
  The second technique compares the difference in mutual information between the independent case and the joint case to describe the success of predicting the dependent series. 
  %The second technique uses {\color{black} mutual information between predictions and observations} to describe the success of predicting the dependent series. 
  When predictability is increased by considering the past values of the driving variable, statistical causality is observed.
  
  \subsection{Linear Causality}
  
  We model a time series as autoregressive by expressing its value $Y_t$ at time $t$ as a sum of the contributions over $m$ distinct lagged series, using the linear equation:

  \begin{eqnarray}
    \label{eq:autoregression1}
    Y_t = \sum_{k=1}^m \beta_{k}^{(Y)} \, Y_{t-k} + \epsilon_t ,
  \end{eqnarray}

  where $\beta^{(Y)}_k$ is a general coefficient term and $\epsilon_t$ is the residual. Linear regression estimates the coefficient parameters $\beta^{(Y)}_k$ which minimise the sum of squared residuals. 

  To detect whether the values of some second time series $X$ anticipate the future values of $Y$, we can compare equation \ref{eq:autoregression1}  with: 

  \begin{eqnarray}
    \label{eq:autoregression2}
  Y_t = \sum_{k=1}^m \beta_{k}^{\prime(Y)} \, Y_{t-k}  \; + 
        \sum_{k=1}^m \beta_{k}^{\prime(X)} \, X_{t-k} 
  + \epsilon^{\prime}_t \;.
  \end{eqnarray}

We determine that the distribution $Y$ is Granger-caused by $X$ if the residual in the second regression is significantly smaller than the residual in the first. When this holds, then there must be some information transfer from $X$ to $Y$. Following Geweke \cite{geweke1984measures}, we can represent the information transfer by:

  \begin{eqnarray}
    \label{eq:geweke_measure}
    \operatorname{TE_{X\rightarrow Y}} = \frac12\log \left(
      \frac{\operatorname{var}(\epsilon_t)}
          {\operatorname{var}(\epsilon^{\prime}_t)}
  \right) ,
  \end{eqnarray}

    where we adopt the transfer entropy notation (TE), following the result from Barnett et al. \cite{barnett2009granger} showing Granger causality to be equivalent to transfer entropy for multivariate normal distributions.

  \subsection{Non-Linear Causality}

  To detect non-linear causality, we apply an information-theoretic approach. Equation \ref{eq:geweke_measure} measures the extent to which the additional information in the lagged variable reduces the variance in the model residuals. Transfer entropy extends this concept by considering the uncertainty, instead of the variance. Adopting Shannon's measure of information \cite{shannon1948}, we can express the uncertainty associated with the random variable $X$ by:

  \begin{eqnarray}
    \label{eq:entropy}
    H(X) = - \sum_{x}  { p(x) \log{p(x)}  } ,
  \end{eqnarray}
  where $H(X)$ is termed the Shannon entropy of the distribution, and $p(x)$ represents the probability of $X = x$.

  This can be conditioned on a second variable to give the conditional entropy:

  \begin{eqnarray}
    \label{eq:conditional_entropy_2D}
    H(Y | X) = H(X,Y) - H(X) .
  \end{eqnarray}

  Where two random variables share information, the mutual information is given by:

  \begin{eqnarray}
    \label{eq:mutual_information}
    I(X;Y) = H(Y) - H(Y | X) .
  \end{eqnarray}
 
  The entropy of $Y$ conditioned on two variables is:

  \begin{eqnarray}
    \label{eq:conditional_entropy_3D}
    H(Y | X, Z) = H(X,Y,Z) - H(X, Z) ,
  \end{eqnarray}
  
  and the conditional mutual information is therefore:

  \begin{eqnarray}
    \label{eq:conditional_mutual_information}
    I(X;Y|Z) = H(Y|Z) - H(Y | X, Z) .
  \end{eqnarray}

  Now, for each lag $k$, we can describe the information transfer from $X_{t-k}$ to $Y_t$ in terms of the following conditional mutual information:

  %\begin{eqnarray}
    \begin{multline}
      \label{eq:TE}
      TE^{(k)}_{\textbf{}X \rightarrow Y} = I(Y_t ;X_{t-k} | Y_{t-k} ) = H(Y_t | Y_{t-k}) \; - \; \\ H(Y_t | X_{t-k}, Y_{t-k} ) \;.
    \end{multline}
  %\end{eqnarray}
 This represents the resolution of uncertainty in predicting $Y$ when considering the past values of both $Y$ and $X$, compared with considering the past values of $Y$ alone. 

  Considering equations \ref{eq:conditional_entropy_2D} and \ref{eq:conditional_entropy_3D}, we can therefore represent the transfer entropy for a single lag $k$, which is shown in equation \ref{eq:TE}, in terms of four separate joint entropy terms. Following equation \ref{eq:entropy}, these may be estimated from the data using a nonparametric density estimation of the probability distributions. For multivariate normal statistics, equations \ref{eq:TE} and \ref{eq:geweke_measure} coincide \cite{barnett2009granger}.

\section{Methods} \label{s.method}

  We calculate linear transfer entropy using ordinary least squares regression, by comparing the variance of the residuals in the joint vector space $ \{Y_t, Y_{t-k}, X_{t-k} \} $ against those in the independent vector space $ \{Y_t, Y_{t-k}\}$, following equation \ref{eq:geweke_measure}. 

  To detect non-linear transfer entropy, we perform nonparametric density estimation to calculate the joint entropy terms in equations \ref{eq:conditional_entropy_2D} and \ref{eq:conditional_entropy_3D}. The density is estimated using a multidimensional histogram approach, where the choice in partitioning of the vector space impacts the calculation of the transfer entropy. In this paper we adopt a partitioning approach which to our knowledge is new in entropy estimation, and which we demonstrate to be robust to varying the coarseness of the partition. Specifically we use a quantile-based binning approach in the marginals, which results in bin edges by each dimension containing equal numbers of data points. 
  
 {\color{black} 
 To partition the sample space in this way, we select each dimension and calculate bin edges independently to contain roughly equal numbers of data points. These are used to construct multi-dimensional histograms for estimating the probability distribution. We observe that the quantile bin sizes perform better than equal-sized bins, as large gradients in the probability distribution function are better able to be captured without the introduction of additional information through refining the partition. }

  In estimating Shannon entropy, the coarseness of the partition directly impacts the {\color{black} numerical value}, with finely-partitioned histograms returning larger entropy values over the same data {\color{black} since more information is acquired about the distribution. This effect should cancel out in the calculation of transfer entropy, however we observe instead that more bins generally results in larger transfer entropies for the same data, which amplifies both signal and noise. We therefore adopt a parsimonious approach in this paper, using a small number of bins compatible with a sufficient resolution, to capture the information transfer. 
  We tested granular partitions of 3 to 8 classes per dimension, finding comparable results in each case. We report the results using histograms of 6 classes per dimension, a partition size which leads to good and meaningful results for each of the currencies analysed.
  }
  
  It is a feature of the nonparametric estimation of entropy that the absolute scale of the transfer entropy measure has only limited meaning; to detect causality, a relative position must be considered. A simple technique proposed by Marschinski \& Kantz \cite{Marschinski2002} is the Effective Transfer Entropy (ETE), derived by subtracting from the observed transfer entropy an average transfer entropy figure calculated over independently-shuffled time series, which destroys the temporal order and hence any possible causality.

  We adopt a shuffling approach producing 50 null-hypothesis transfer entropy values from independently shuffled time-series over the same domain, containing no causality. By calculating the mean and standard deviation of the shuffled transfer entropy figures, we estimate the significance of a causal result as the distance between the result and the average shuffled result, standardising by the shuffled standard deviation:
  
  \begin{eqnarray}
    \label{eq:Z-score}
    Z := \frac{\operatorname{TE} - \bar{\operatorname{TE}}_{\operatorname{shuffle}}}
              {\sigma_{\operatorname{shufle}}} .
  \end{eqnarray}

  This corresponds to the degree to which the result lies in the right tail of the distribution of the zero-causality { shuffled samples}, and hence how unlikely the result is due to chance. Therefore the Z-score figure represents the significance of the excess transfer entropy in the un-shuffled case. We compute the Z-score in Eq.\ref{eq:Z-score} for both linear and non-linear results.

  To justify the usage of these techniques in detecting causal relationships in practice, we first validate the methodology using coupled time series of predefined causative relationships. 

  %\section{\label{sec:level2} Synthetic Geometric Brownian Motion}
  \subsection{Synthetic Geometric Brownian Motion}
  We validate the approach by generating synthetic data following a directionally coupled random walk. First, we generate a driving series, following a discrete Geometric Brownian Motion (GBM):

  \begin{eqnarray}
      \label{eq:GBM}
      X_{t+1} = (1+\mu) X_t + \sigma X_t \; \eta_t ,
  \end{eqnarray}

  %  \begin{eqnarray}
  %    \label{eq:GBM}
  %    dX_t = \mu X_t dt + \sigma X_t dW ,
  %  \end{eqnarray} 
  where $\eta_t$ is a normally distributed random noise  $\eta_t \sim \mathcal{N}(0,1)$, and $\mu$ and $\sigma$ are respectively drift and diffusion coefficients. {\color{black} Then we produce a dependent series $Y_t$, which is a linear combination of $X$ and a second, independent GBM process $X^{\prime}$, the strength of the dependency being determined by some coupling constant $\alpha$, over some lag length $k$:

  \begin{eqnarray}
    \label{eq:coupled_random_walk}
    Y_t = (1-\alpha) X_{t-k}  + \alpha X^{\prime}_{t-k} .
  \end{eqnarray}

  }
  %\section{\label{sec:level2} Synthetic Coupled Logistic Map}
  \subsection{Synthetic Coupled Logistic Map}
  We generate non-linear coupled time series using a coupled logistic map. This system can be represented in terms of two stationary difference equations; the independent series is defined by the difference equation given by the general update function $f(X)$:

  \begin{eqnarray}
    f(X_t) = X_{t+1} = rX_t(1-X_t)    
    \label{eq:f(x)}
  \end{eqnarray}

  where $X_t$ is the value of $X$ at time $t$, and $r$ is a parameter which in fact defines the dynamical state of the system. {\color{black}Following Hahs \& Pethel \cite{hahs2011distinguishing}, we take $r=4$ so the function evolves chaotically.} We then introduce a second map, which is dependent on the first, taking the form:

  \begin{eqnarray}
    Y_{t+1} = (1-\alpha) rY_t(1-Y_t)  + \alpha g(X_t)   
    \label{eq:coupled_map}
  \end{eqnarray}

  where $ \alpha \in [0,1]$ is the cross-similarity, or coupling strength, and $g(x)$ is a coupling function which may be chosen to produce different dynamic effects. We follow the choice of Boba et al. \cite{boba2015efficient} and Hahs \& Pethel \cite{hahs2011distinguishing} in the coupling function:

  \begin{eqnarray}
    g(X_t) = (1-\epsilon) f(X_t) + \epsilon f(f(X_t)) 
    \label{eq:g(x)}
  \end{eqnarray}

  where $\epsilon \in [0,1]$ represents the coupling strength, describing the extent to which $Y_{t+1}$ depends on $f(f(X_t))$. It should be noted that the logistic map, in contrast to geometric brownian motion, is a deterministic, albeit chaotic system, and that therefore $f(f(X_t))$ is equivalent to $X_{t+2}$. {\color{black} The extent of this anticipatory effect is driven by the selection of the $\epsilon$ parameter. We follow Hahs \& Pethel in selecting $\epsilon=0.4$. Indeed, as $\alpha$ increases, with large $\epsilon$, the direction of information transfer is less clear, as $Y_t$ contains more information about the future values of $X$. }

\section{Validation with Synthetic Data} \label{s.validation}

  In order to validate the autoregressive and information-theoretic approaches to detecting causality, we apply these to the calculation of transfer entropy for synthetic data generated by both linear and non-linear coupled time series, of increasing coupling strength.

  \subsection{Linear Process Causality Validation} 

  We calculate the directional information transfer from the driving series to the dependent series, and in the reverse direction, using both autoregressive and information-theoretic approaches for the linearly-coupled system of GBM walks defined by equations \ref{eq:GBM} and \ref{eq:coupled_random_walk}. Figure \ref{fig:GBM_confirmation} shows the results for coupling strengths from $\alpha = 0.0$ to $\alpha = 0.5$. For each coupling strength, a data set is simulated over 2500 time steps. Both techniques are applied to each data set to calculate the information transfer, in both directions, with the results from $X \rightarrow Y$ and from $Y \rightarrow X$ plotted on separate axes. 

  In the information-theoretic approach we calculate transfer entropy using histograms with quantile binning of 6 classes per dimension. We generate multiple synthetic coupled random walks, calculating transfer entropy and Z-scores for each realisation, and reporting the mean values. Quantile bins are generated independently for each realisation.

  We observe that {\color{black} using finer-grained partitions, hence of more bins, results in an increased estimate of transfer entropy for the same data. However, the choice of coarseness does not affect the final analysis in validating causality; equivalent results are observed when considering significance instead of just the numerical transfer entropy figure.}

  As can be observed from Fig.\ref{fig:GBM_confirmation}, the qualitative correspondence between both methods is clearly visible, and quantitatively the results are similar. Additionally, the one-way direction of information transfer is accurately detected, with large transfer entropy and Z-scores observed in the direction of $X \rightarrow Y$, and small values in the opposite direction.

  %\begin{figure*}[!htb]
    
  %\end{figure*}

  \subsection{Non-Linear Process Causality Validation}

  We calculate the directional information transfer from the driving series to the dependent series, and in the reverse direction, using both autoregressive and information-theoretic approaches for the non-linear coupled logistic map system from equations \ref{eq:f(x)}, \ref{eq:coupled_map} and \ref{eq:g(x)}.  

  {\color{black}
  In the information-theoretic approach we {\color{black} calculate transfer entropy again using histograms with quantile binning of 6 classes per dimension, generating bins independently for each realisation.}}
  
\begin{figure*}[!htb]
    \includegraphics[width=\linewidth]{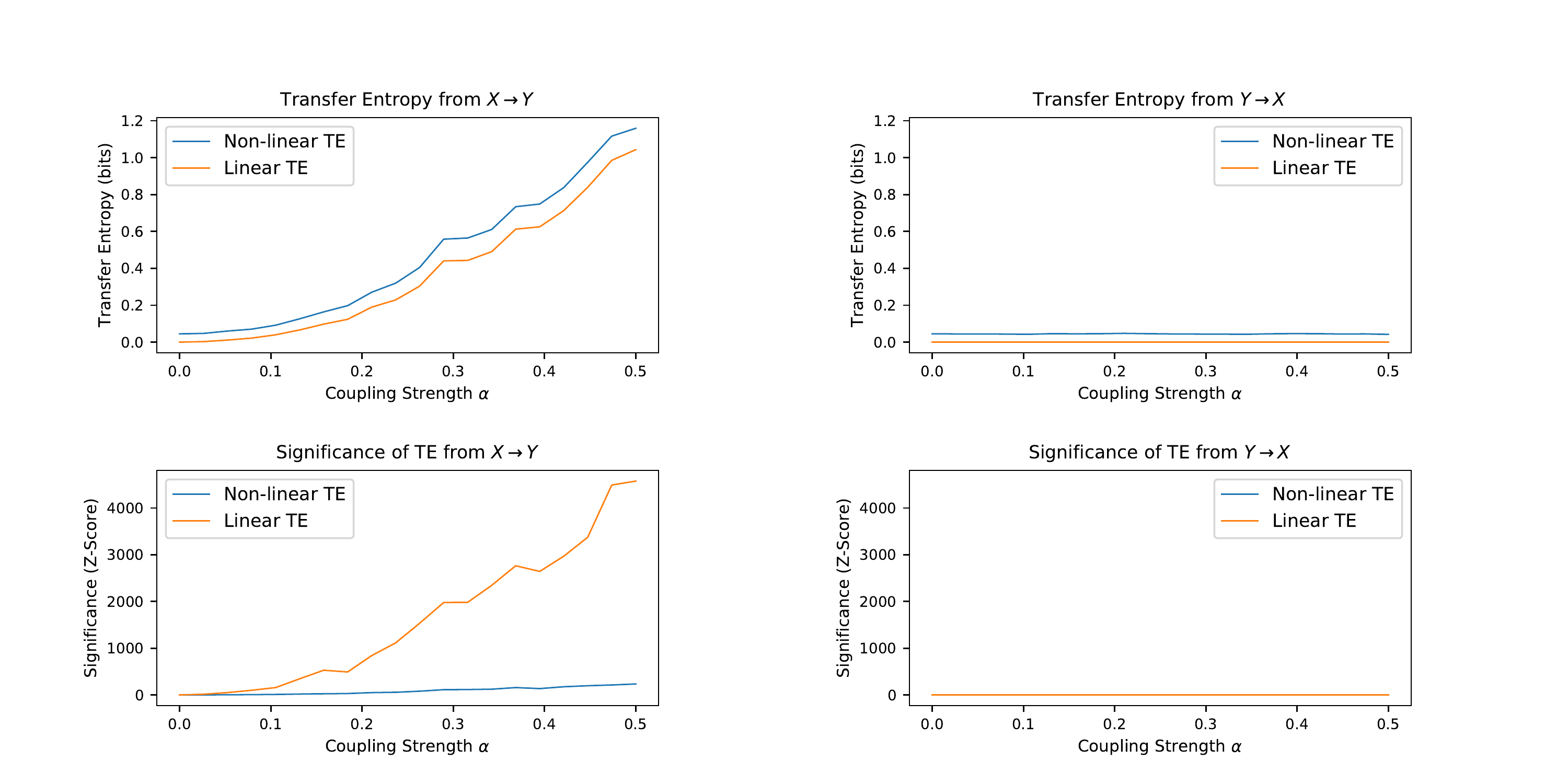}
    \caption{Demonstration that both linear and non-linear transfer entropy methods detect causality for linearly coupled synthetic data. The plots are calculated over 2500 data points of the synthetic random walk process from equations \ref{eq:GBM} and \ref{eq:coupled_random_walk}. Non-linear transfer entropy is calculated using a quantile histogram of 6 classes per dimension. The Z-score of each result is also plotted for both methods. {\color{black} We observe a small but non-zero baseline transfer entropy in the non-causal direction $Y \rightarrow X$, which explains the systematic over-estimation of transfer entropy calculated in the direction $X \rightarrow Y$. The size of this over-estimation increases with the number of histogram bins.}}
    \label{fig:GBM_confirmation}

    \includegraphics[width=\linewidth]{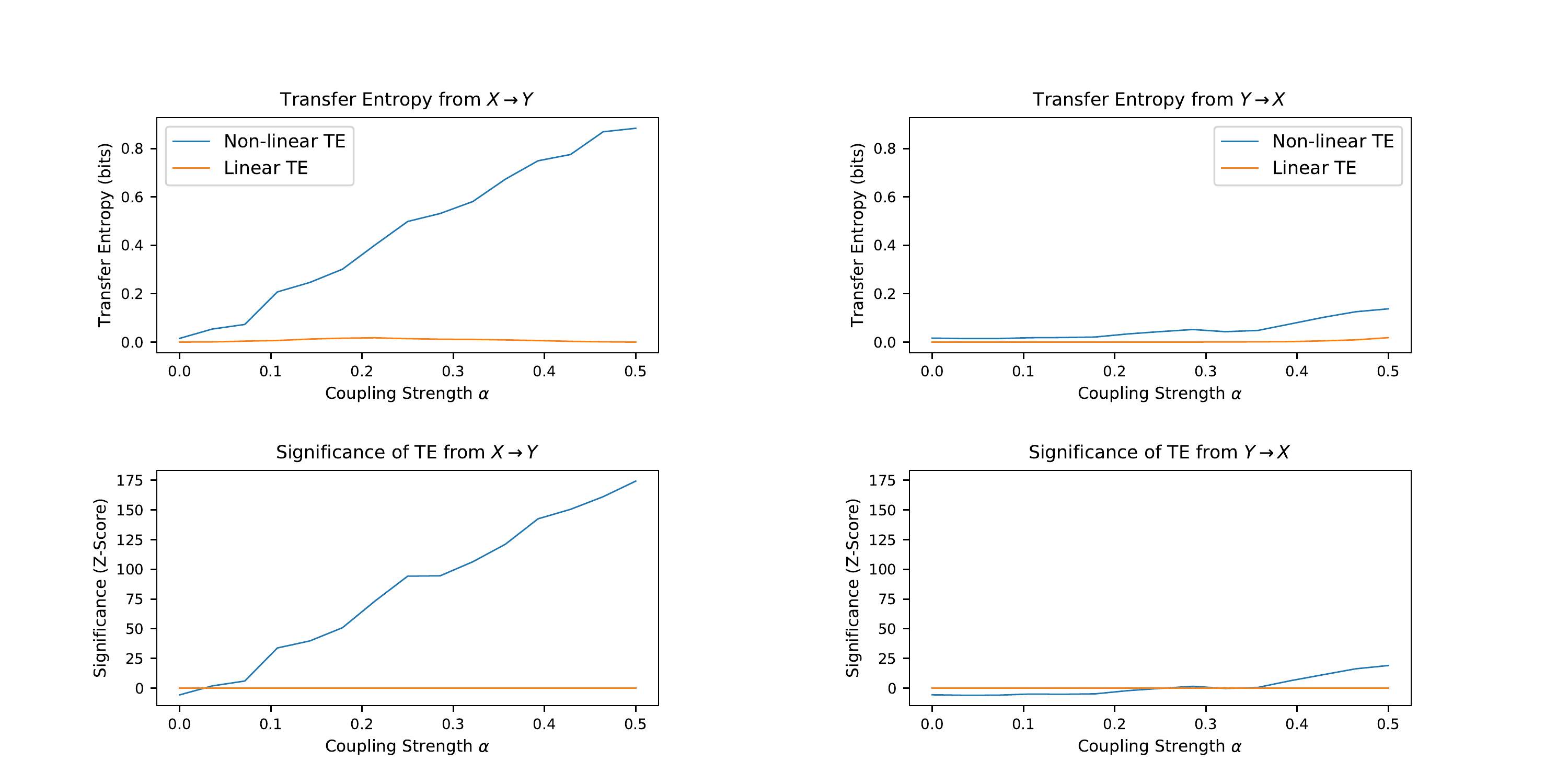}
    \caption{Demonstration that the non-linear causal relationship in synthetic data generated from equations \ref{eq:coupled_map} and \ref{eq:g(x)} is detected only by the non-linear method. The plots are calculated over 2500 data points of the synthetic coupled logistic map process, with $\epsilon = 0.4$. Non-linear transfer entropy is calculated using a quantile histogram of 6 classes per dimension. The Z-score for each result is also plotted for both methods. {\color{black} We note that from $\alpha=0,5$ there is some detection of information transfer in the other direction, using both methods; this is observed to increase as $\alpha$ approaches $1$.}
    }
    \label{fig:CLM_confirmation}
  \end{figure*}

  Figure \ref{fig:CLM_confirmation} shows the mean transfer entropy results for 2500 synthetic data points. We observe that, for this system, the linear method is incapable of detecting causality; it finds no significant information transfer, fails to represent the expected exposure–response relationship and also suggests a slight causality in the reverse direction. The information-theoretic method, by contrast, produces results which better represent the increasing coupling strength relationship, and direction of causality in the system. However, this technique also detects causality from $Y$ to $X$, for large values of $\alpha$, and the effect is greater than in the linear case. We explain this with reference to the coupling function $g(x)$ which involves repeated application of the update function $f(x)$; from equation \ref{eq:f(x)} we see that $f(f(X_t))$ is equivalent to $X_{t+2}$ so, for large $\alpha$, $Y_t$ will contain increasing amounts of the future information of $X_t$. {\color{black} In fact, at large coupling strengths approaching $\alpha=1$, the observed transfer entropy from $X$ to $Y$ begins to decrease, as more information exists in $Y$ about its future evolution.}

  The results of these validation experiments suggest that the information-theoretic approach is superior in detecting causal signals, being model-free and so able to detect relationships of more complex, non-linear modes. 

  \subsection{Decay of Causal Signals with Lag Length}
  As a final validation exercise, we explore the performance of the {\color{black} methods} in detecting signals in coupled time series when the lag of the relationship is unknown. In general, it is expected that causal links should be strongest at time-lags closest to the true signal lag, and gradually decay as the time-lag considered is increased. However, the complexity of causative relationships, particularly where any feedback exists between the time series, suggests that there could also be multi-modal causalities, operating at different lags.  

\begin{figure*}[!htb]
    \includegraphics[width=\linewidth]{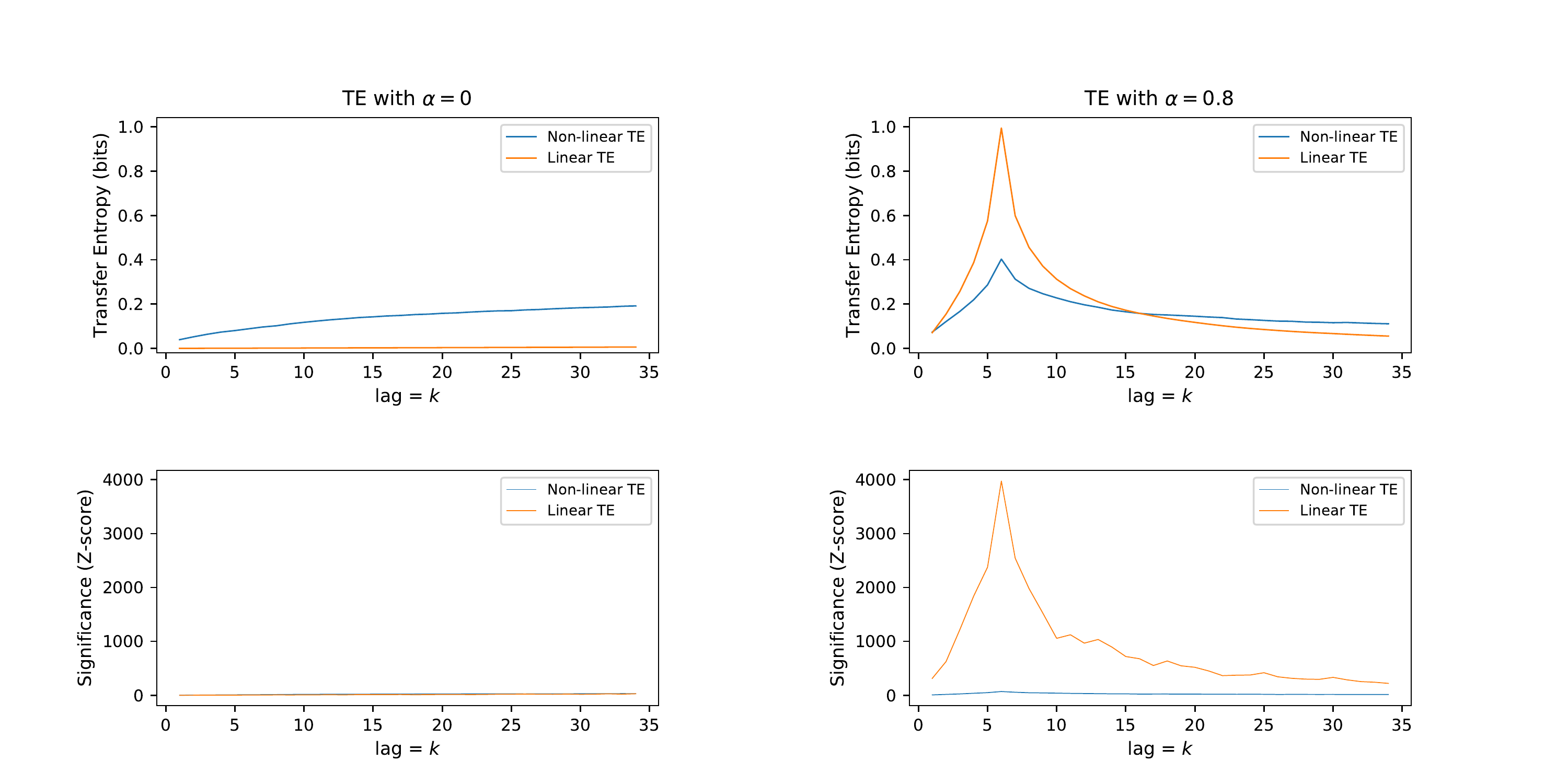}
    \caption{ Demonstration that both methods identify the true lag $L=6$  with maximal transfer entropy. Non-linear transfer entropy is calculated using a quantile-binned histogram, of 6 classes per dimension, over 2500 points. The Z-score for each result is also plotted for both methods. {\color{black} We observe a non-zero transfer entropy in the non-causal case $\alpha = 0$, which grows with time-lag $k$. This might explain the systematic over-estimation of transfer entropy calculated in the direction $X \rightarrow Y$. The size of this over-estimation increases with the number of histogram bins.}
    }
    \label{fig:dropoff}
  \end{figure*}

  We use the coupled GBM system defined in equations \ref{eq:GBM} and \ref{eq:coupled_random_walk} to create a coupling of a fixed lag $L=6$, and then perform both autoregressive and information-theoretic analysis to detect the transfer entropy at time-lags from $k=1$ up to $k=35$. {\color{black}
  The information-theoretic approach is again applied using histograms partitioned into 6 classes per dimension. }
  The results are shown in Fig. \ref{fig:dropoff}.

  We observe two interesting features. First, the surprising anticipation of the peak is seen at lags $k$ shorter than the true lag $L$ of the causal relationship. Secondly, a clear peak is seen at the expected lag, which decays slowly and incompletely. We explain this by the comparison to the transfer entropy observed in the decoupled case with $\alpha=0$. In the limit of increasing time-lag $k$, the information-theoretic approach detects a causality even when there is no coupling in the data; we note that the Effective Transfer Entropy measure could perform better in such cases, where subtracting the average zero-causality transfer entropy would give a better estimate of the true information transfer \cite{Marschinski2002}. Importantly, both techniques show a clear peak at the true causal time-lag, with the autoregressive technique displaying considerably greater significance, albeit this is also observed even at spurious lags. It is possible that the observed trend of increasing causality at long lags is due to the way in which data points are excluded for increased lags; for $k=35$, for example, we discard $35$ data points from the set in the calculation of transfer entropy.

%\clearpage
\section{Results with Real Data} \label{s.results}
  Having confirmed that the information-theoretic approach is able to detect both linear and non-linear signals, we apply the technique to investigate the effect of social media sentiment on cryptocurrency prices. We also apply the linear method to compare whether linear or non-linear dynamics dominate any causal relationship.

  We estimate information transfer over 24-month windows, rolling forward with a stride of two weeks from the earliest market data available to September 2018. Price is taken as the combined close price, on the hour, over an aggregation of exchanges (see appendix \ref{a.data}). Social sentiment is estimated from NLP analysis of Twitter tweets and StockTwits during the preceding hour; we quantify this sentiment as the sum of of positive messages in the previous hour. In early periods of the data, infrequently some hours have no messages; in these cases we forward-fill from the previous hour, making the assumption that sentiment does not drop to neutral in these cases. To handle non-stationarity in the data, we take the difference between the logarithms of the values at times $t$ and $t-1$. This differencing is applied to both time series. 

  The choice of timescale in aggregating raw sentiment data involves a trade-off; with too fine a timescale, there are not enough messages to estimate sentiment, but too long a timescale cannot capture the dynamics of the time series. We hypothesise that causal signals between sentiment and price operate at sub-hourly timescales; hourly aggregation is the smallest time period available in the data, and so this aggregation of sentiment is used. 

  The transfer entropy is calculated over multiple backward-looking 24-month windows, which are passed over all available data with a two-week stride. For the information-theoretic approach, {\color{black} it is observed that performing the analysis with histograms of equal-width bins} gives different results depending on the number of bins selected. {\color{black} Specifically, partitioning the axes of the sample space into odd-numbers of bins produces no significant result over this data, suggesting the information is captured mostly from the middle peak of the distribution. However, we note that the use of quantile binning avoids this issue, finding both odd-numbered and even-numbered bin counts to provide similar results, suggesting a key benefit in using quantile bins for the calculation of transfer entropy.} Accordingly, in this analysis we partition the sample space into quantile bins, using six classes per dimension, having validated this choice in Section \ref{s.validation}.
  
  The histogram bins for the non-linear approach are calculated once, using the full data set for each currency, and then they are applied across all windows. In selecting an appropriate partition, further bias is inevitably introduced. By calculating appropriate bins for each window, the results cannot be directly compared between windows. {\color{black} However, the growth in message volumes over time means that selecting bins sized to capture the full spread of values also introduces a bias, since such bins are  more suited to the later months than the earlier months. Additionally, since} the granularity of the histogram partition also impacts the transfer entropy value, {\color{black} we perform significance tests over each window independently, to report any causality, calculating the Z-scores and comparing these across windows and currencies.} 

  We report the windows with greatest significance using a time-lag of $k=1\operatorname{hour}$. Performing the analysis using longer time-lags shows weaker causal signals over this data. This provides evidence in support of the hypothesis that the true causal dynamics operate at sub-hourly timescales.
   
  {\color{black} We report results for linear and non-linear transfer entropy, calculated using multidimensional histograms using bins of 6 quantile classes per dimension. The transfer entropy figure and Z-score are calculated independently for each 24-month window, bins are generated once for each currency, over the whole dataset, and used for each window. This selection produces the clearest detection of a causal relationship between sentiment and price.}

\begin{figure*}[!htb]
    \includegraphics[width=\linewidth]{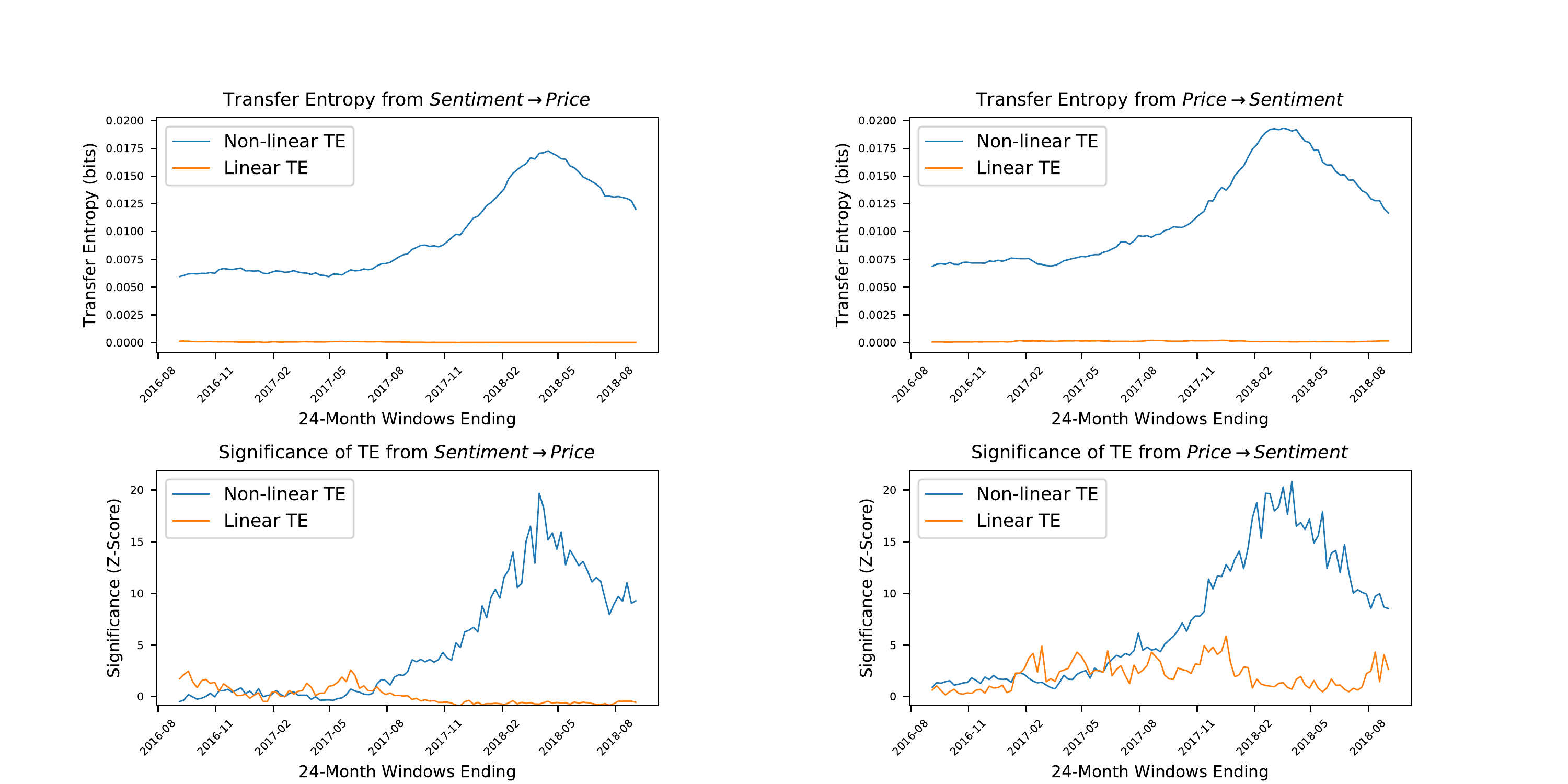}
    \caption{Evidence that BTC sentiment and price are causally coupled in both directions in a non-linear way. Non-linear TE is calculated by multidimensional histograms with 6 quantile bins per dimension. Z-scores, calculated over 50 shuffles, show a high level of significance, especially during 2017 and 2018, in both directions.}
    \label{fig:BTC_TE}
  
    \includegraphics[width=\linewidth]{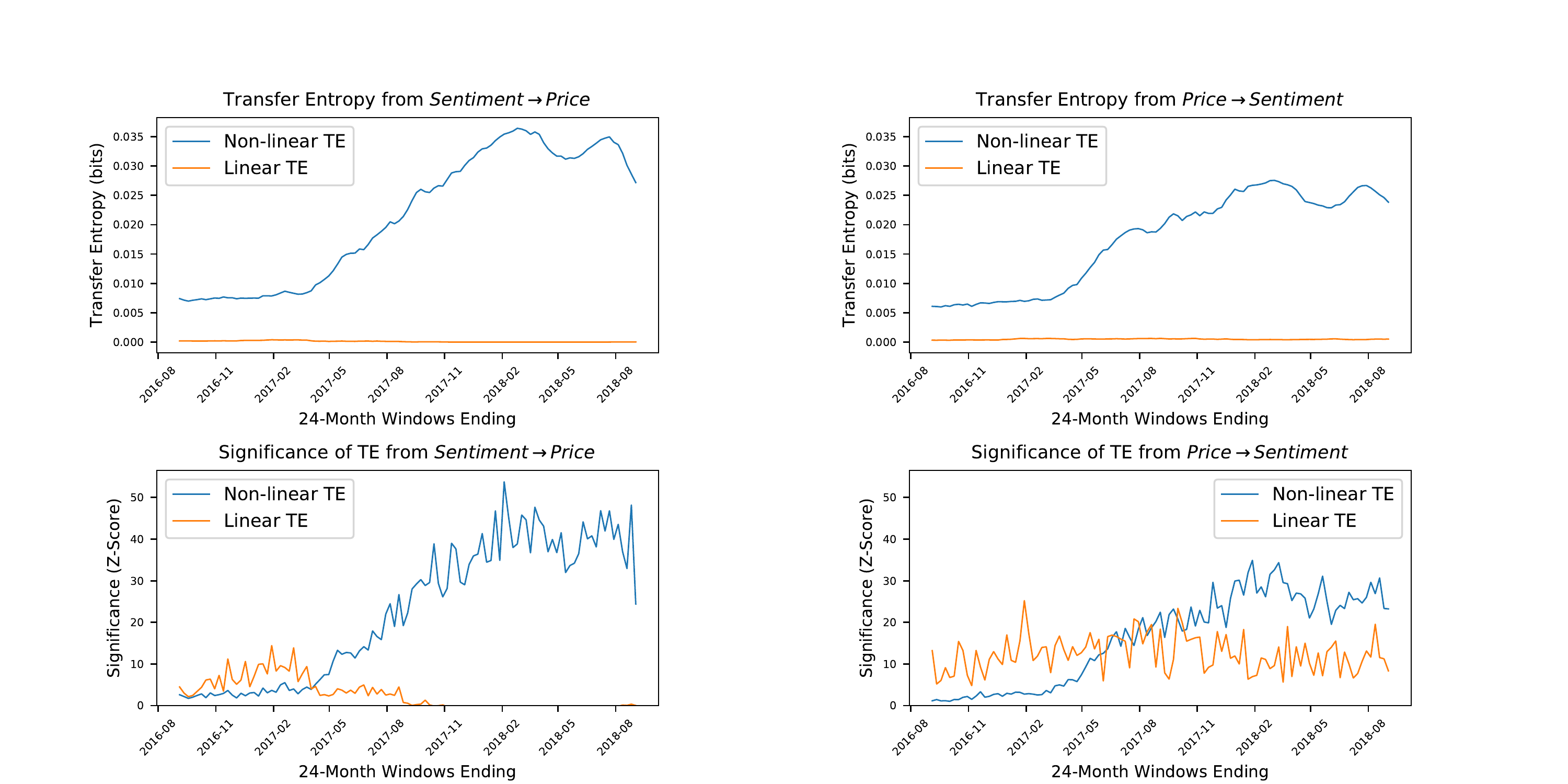} 
    \caption{Evidence that LTC price and sentiment are causally coupled in both directions in a non-linear way, with sentiment having a larger influence on price than the other way round. Non-linear TE is calculated by multidimensional histograms with 6 quantile bins per dimension. Z-scores, calculated over 50 shuffles, show a small but clear significant signal, in both directions, with the net information transfer generally operating in the direction of sentiment to price.}
    \label{fig:LTC_TE}
  \end{figure*}

\begin{figure*}[!htb]
    \includegraphics[width=\linewidth]{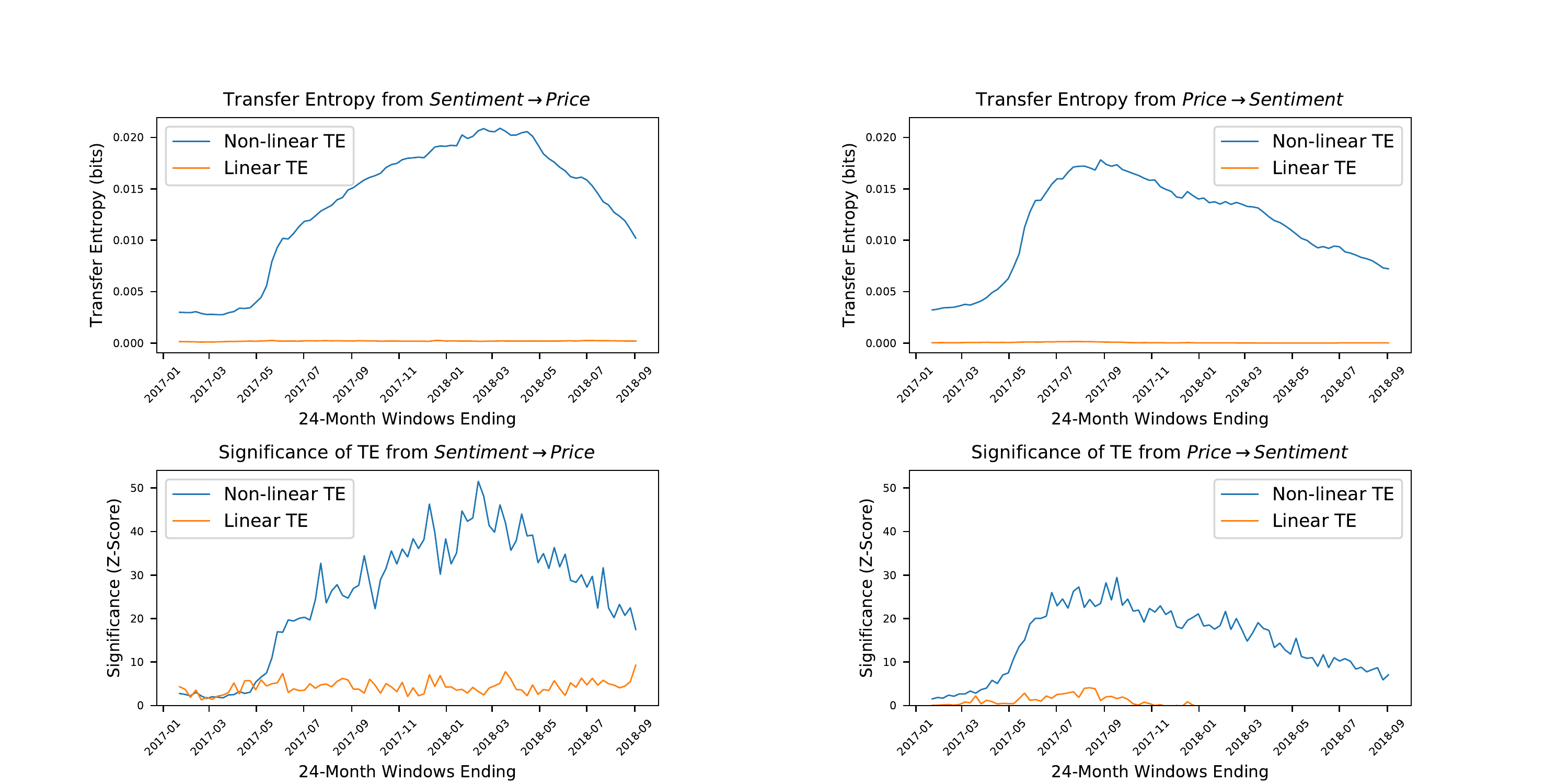}
    \caption{Evidence that XRP price and sentiment are causally coupled in both directions in a non-linear way, with the prevailing direction of information transfer flowing from sentiment to price in the first period, and from price to sentiment in the second. Non-linear TE is calculated by multidimensional histograms with 6 quantile bins per dimension. Z-scores, calculated over 50 shuffles, show a small but clear significant signal, in both directions, which decays rapidly towards January 2018 and does not recover afterward. 
%{\color{black}\bf specify number of bins}                   
}
    \label{fig:XRP_TE}
  \end{figure*}

\begin{figure*}[!htb]
    \includegraphics[width=\linewidth]{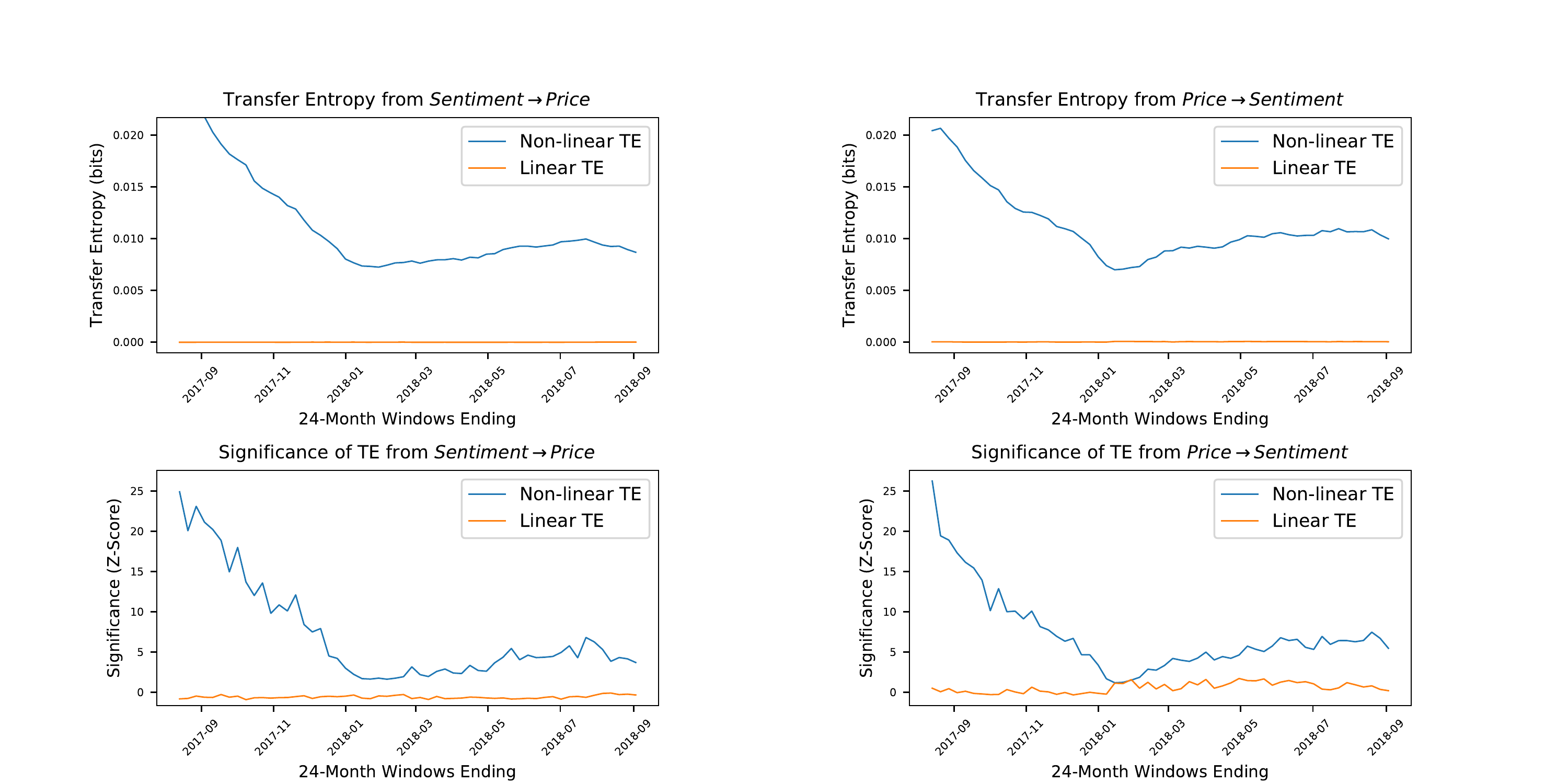}
    \caption{Evidence that ETH price and sentiment are causally coupled in both directions in a non-linear way. Overall this coupling is of lower significance compared to the other currencies investigated. Non-linear TE is calculated by multidimensional histograms with 6 quantile bins per dimension. Z-scores,  calculated over 50 shuffles, indicate some significance, followed by low significance after the collapse in signal strength beginning around January 2016.        
  }
    \label{fig:ETH_TE}
  \end{figure*}

  Plots showing the information transfer for the four cryptocurrencies investigated are reported in Fig. \ref{fig:BTC_TE}, Fig. \ref{fig:LTC_TE}, Fig. \ref{fig:XRP_TE} and Fig. \ref{fig:ETH_TE}.

  For BTC, in Fig. \ref{fig:BTC_TE}, we detect a strong causative signal, of roughly similar scale in both directions of sentiment to BTC price and in the reverse direction.

%\end{figure*}

LTC, in Fig. \ref{fig:LTC_TE}, shows a similar pattern to BTC, although it is less equivocal in the direction of information transfer, with significance in the direction of sentiment to price consistently appearing greater than in the reverse direction. We note the Z-scores reveal greater overall significance compared to the other currencies. 

%begin{figure*}

  XRP, in Fig. \ref{fig:XRP_TE}, shows a clear non-linear causality from sentiment to price and also in the opposite direction. However, the signal is more significant from sentiment to price, and especially in the periods ending in 2018. 
  
  ETH, in Fig. \ref{fig:ETH_TE}, shows an interesting and unique behaviour. In particular, there appears to be, initially, a significant signal which collapses in both directions in the windows ending around January 2018. This strongly suggests another driving mechanism, the effect of which first becomes present around January 2016 (due to 24-month windows). This effect is likely to be associated with the rapid price movements at the time. %\cite{Ethereum}
  
\section{Conclusion} \label{s.conclusions}

Information-theoretic and autoregressive techniques were developed and validated on coupled random walks and chaotic logistic maps, confirming the ability of both techniques to detect linear information transfer, and of the information-theoretic technique to detect non-linear information transfer. Following validation, the techniques were applied to historical data describing social media sentiment and cryptocurrency prices to detect information transfer between sentiment and price movements. 

  The information-theoretic investigation detected a significant non-linear causal relationship in BTC, LTC and XRP, over multiple timescales and in both the directions sentiment to price and price to sentiment. The effect was strongest and most consistent for BTC and LTC.  Given the hypothesis that low barriers to entry and unsophisticated investor speculation are key drivers for price movements, {\color{black} and that these represent the most widely known and traded cryptocurrencies, the fact that causality is detected most clearly for these currencies corresponded to expectations. We observe that the direction of information transfer is stronger from sentiment to price for all currencies except BTC, for which the causal signal is slightly stronger in the direction from price to sentiment.}

  The significance tests confirm the existence of {\color{black} causally-coupled relationships}, though the strength of these are challenging to accurately quantify from the data, especially for the sake of comparison between different time series and between the linear and non-linear results over the same data. However, the significance values themselves offer the possibility of quantifying the strength of causality, which may be used as a proxy when {\color{black} using transfer entropy as a tool for detecting statistical causality}. With this work we demonstrate that the dynamics of the causative relationship is non-linear, as the autoregressive technique observed {\color{black} at most very limited causality} in either direction, for any of the currencies.

  Let us point out that there is a risk of assuming ergodicity in the results; we have shown the level of causation in-sample, but there is no fundamental reason that this should continue out-of-sample. Up to this point, research into information transfer has been restricted to backwards-looking statistical analyses, overlooking any analysis into the forward evolution of causal relationships with time.

%\clearpage

\section{Acknowledgments}
  The authors acknowledge Th\'arsis Souza in advising on the method of testing for linear Granger causality, with thanks along with Yuqing Long, whose data collation and wrangling was a great help. Finally, a great debt of thanks is to PsychSignal for providing their market sentiment data for this academic study.  
 TA acknowledges support from ESRC (ES/K002309/1),  EPSRC (EP/P031730/1) and EC (H2020-ICT-2018-2 825215).
  \newline

\appendix

\section{Appendix}

  \subsection{Source Code}
    All analysis for this paper was performed using a Python package (PyCausality) created during the lead author's MSc. This is maintained on the author's public GitHub profile, which can be found at \url{https://github.com/ZacKeskin/PyCausality}. For the latest release this can be simply installed via PyPi using pip.

    Ongoing maintenance and pre-release development of the package will be made available through this repository, and contributors may fork code and submit pull requests to develop this further.

  \subsection{Data} \label{a.data}

  The social sentiment data was provided courtesy of PsychSignal, and may be made available pending request to the authors. The data takes the form of the number of positive messages and the number of negative messages, publicly shared on either Twitter or StockTwits, associated each hour with the cryptocurrencies in question. The association is detected via the use of a `hashtag' (or `cashtag') which takes the form of \#BTC or \#Bitcoin (for example) on twitter, or \$BTC on StockTwits.  For inclusion in the dataset, the message must contain one of the tags described in Table \ref{tab:table1}.  
  
  Price data is the hourly close in USD, obtained via CryptoCompare's public API. This provides a combined average over multiple exchanges, where prices are available. For further details, the documentation is available at \url{https://min-api.cryptocompare.com/}

  \begin{table*}[!htb]
    \caption{\label{tab:table1} Hashtags used to map social media messages to specific cryptocurrencies.}
    \begin{ruledtabular}
      \begin{tabular}{cccc}
      %\begin{tabular}{|| m{5em} | m{10em}|| m{5em} | m{10em}||} 
        Curency & Tag & Curency & Tag        \\ [0.5ex] 
        \hline 
        Bitcoin & BTC &  Litecoin & LTC.X    \\
        Bitcoin & BCOIN & Litecoin & LTCUSD  \\
        Bitcoin & BTC.X & Ripple & XRP.X     \\
        Bitcoin & BTCEUR & Ripple & XRPBTC   \\
        Bitcoin & BTCGBP & Ripple & XRPUSD   \\
        Bitcoin & BTCUSD & Ethereum & ETH    \\
        Bitcoin & GBTC & Ethereum & ETH.X    \\\
        Bitcoin & SGDBTC & Ethereum & ETHUSD \\
      \end{tabular}
    \end{ruledtabular}
    \end{table*}

%\textit{Physical Review} style requires that the initial citation of
%figures or tables be in numerical order in text, so don't cite
%Fig.~\ref{fig:wide} until Fig.~\ref{fig:epsart} has been cited.
%\input acknowledgement.tex   % input acknowledgement

%\bibliography{references.bib}
%\nocite{lizier}
%\nocite{boba2015efficient} 
%\nocite{he2017comparison}
%\nocite{lutkepohl2005new}
%\nocite{san2014unraveling}
%\nocite{wilmott2007paul}
%\nocite{scott2015multivariate}
%\nocite{reshef2011MIC}
%\nocite{scott1979optimal}
%\nocite{knuth2006optimal}
%\nocite{coverandthomas}

\end{document}